# Wi-Fi Scaling and Performance in Dense Frequency Planned Networks

Jonathan Ling, Howard Huang, and Reinaldo Valenzuela

ABSTRACT

The area capacity scaling of Wi-Fi has been thought to be poor due to the 802.11 MAC enforcing single transmissions in overlapped cell coverage areas. This is based on the assumption that the energy detection and virtual carrier sense MAC mechanisms at every station correctly detect ongoing transmissions and thus prevent new transmissions. However overlapped or faded packet headers can cause stations in different cells to lose synchronization with each other. When the overlapped transmissions are in different cells the spatial frequency reuse may increase the area capacity. Area capacity improvement due cell densification is investigated numerically assuming a large indoor frequency planned network with fading and packet capture. For comparison, Wi-Fi and LTE base stations are placed at the same cell centers. LTE area capacity grows faster as it is only constrained by SINR. Wi-Fi area capacity grows slower limited by decreasing SINR and the partially operational virtual carrier sense.

## I. INTRODUCTION

Cell densification has been meeting the increasing network demand, but especially when cells overlap, the media access control mechanism (MAC) plays a strong role in determining capacity of the network. Operating in the unlicensed band Wi-Fi networks must be compatible with wide range of devices that can share the band, including ad-hoc deployments of the same technology. The 802.11 standard [1][2] specifies collision sense multiple access (CSMA) MAC, which firstly listens before transmitting, and if a collision occurs, during the next sensing opportunity, it delays its next attempt. There are two carrier sense mechanisms: energy detection (ED) which compares the power received to an ED threshold and virtual carrier sense (VCS) that relies on length information in the packet header. The length field is used to update the network allocation vector (NAV), a timeline of when the channel is busy. Both the ED and VCS prevent collisions by preventing additional transmissions. Two Wi-Fi access points (APs) or clients (STAs) can in fact be put physically on top of each other and timeshare the same channel with minimal amount of collisions.

Notably CSMA is optimal in the sense it maximizes spatial reuse, producing maximal packing patterns, whereby no additional transmissions can be accommodated, e.g. see [3], and is useful in guaranteeing minimum signal-to-noise-interference ratio (SINR). Depending on the topology, network throughput may improve when transmissions are also permitted to overlap and interfere with each other. For cellular technologies due to planned deployment forcing reasonable inter-cell distances, full spatial reuse, yields the best performance. Nonetheless as out-of-cell interference becomes stronger due to the smaller cells the gain due to cell splitting along with capacity-to-cost ratio diminishes [4][5][6]. Moreover for 802.11 networks the CSMA based MAC appears to impose a hard limit on the area capacity due to strict enforcement of timesharing.

However this MAC enforced limit is based on idealistic modeling of the receiver where all stations are connected and coordinated, i.e. always hear each other's transmissions. In practice MAC coordination *between cells* is limited due to propagation loss, fading, and interference. This is demonstrated by the experiment in [7] which shows the throughput growing with separation of two pairs of stations (STAs) due to a gradual transition between a coordinated and an uncoordinated MAC.

We investigate the area capacity scaling of a large indoor Wi-Fi network. To properly model the CSMA MAC, a physical layer packet capture was introduced into NS-3. Packet capture occurs when two or more packets are transmitted simultaneously and at each receiver the strongest packet is decoded. Therefore according to the VCS each receiver may have a different idea when the channel is free or not.

Basic experiments with small number of APs are performed, insights gleaned, followed by experiments on a large network with wrap around propagation. We also compare Wi-Fi performance to 3GPP LTE performance to separate the effect of worsening signal-to-interference-and-noise-ratio (SINR) from MAC interaction. Simulation traces from NS-3 are examined carefully, and from this analysis we provide an explanation of the underlying MAC behavior justifying the observed trends in area capacity.

The performance of 802.11 is difficult to characterize due to the complexity of the MAC, especially when considering 802.11n/ac enhancements, and large number of variables, such as traffic type and loading, wireless channel, and spatial distribution of users. Tools such as open source simulators have difficulty juggling conflicting requirements such as managing complexity while providing useful and correct abstractions from application layer down to the physical layer. NS-3 has 802.11a support and the basic CSMA implementation, but advanced PHY/MAC layer techniques in 802.11n or 802.11ac are not yet supported. Some features in the simulation model lack validation or have taken a while to correct [8][9]. Alternatively experimental approaches have their own difficulties. Some of the MAC, e.g. the scheduler

Jonathan Ling (Jonathan.Ling@nokia.com) is at Nokia.

on the AP, is proprietary and undocumented. Therefore the underlying mechanisms can be difficult to isolate and generalize beyond the specific set of hardware and software. One must rely on low level probing tools that have their own limitations. For example the radiotap header [10] would seem ideal to obtain low level PHY/MAC information, but device drivers tend to fill in fields sparsely and not always correctly.

The authors in [7] point out that researchers must go beyond simplistic (0 or 1) adjacency graphs to model connectivity. Nonetheless it has been a useful starting point taken in analytical work of Bianchi [11] where throughput vs. number of users is determined for a single cell, and multi-cell performance estimates in [12][13]. More accurately packet capture causes the stronger received packet to be decoded, and intuitively this leads to imbalance in user rates in favor of users closer to the AP. In [8] an analytical model for the effect of capture unfairness has been derived and experimentally validated. One solution is to route uplink IP packets on LTE and downlink on Wi-Fi [19], thus both uplink and downlink will be scheduled and fairness enforced. Desensitizing the VCS to avoid time sharing with co-channel cells has been proposed [14] and NS-3 simulations show that to 85% increase in rates after doubling the number of APs in a stadium environment [15]. Other improvements can be found in 802.11ax draft [16].

The remainder of this paper is organized as follows. In Section II we discuss firstly the effect of packet capture on the MAC and secondly modifications to NS-3 model to support capture. In Section III we study the MAC behavior and performance for a small isolated network. In Section IV we study performance of a large frequency planned Wi-Fi network as the cell density increases. In the conclusions we summarize the key discoveries and mention future work.

## II. CARRIER SENSE AND MODELING

### A. Overlapped packets and VCS/ED

VCS is the primary sensing mechanism. It involves (i) physical layer packet preamble detection (ii) channel estimation (iii) header payload decoding and parity check. For normal data frames the channel is declared busy for the duration of the payload plus the short-inter-frame-space (SIFS) and acknowledgement (ACK). Since the packet header is coded at the lowest and most reliable modulation-and-coding-scheme (MCS), detection can occur at the lowest signal strength, approximately -90 dBm with modern receivers. This means that STAs will refrain from using the channel if it detects another transmission, even a very weak one.

Regulatory agencies require power or energy detection in unlicensed bands before transmission, a.k.a. listen-before-talk. According to the 802.11 standard [1] the threshold is specified at -62 dBm which is 20 dB higher than the nominal decoding sensitivity of the lowest MCS (i.e. rate ½ BPSK). The ED threshold is relatively high SNR of 32 dB given 7 dB NF. Energy detection also has an important place in preventing in-cell collisions.

In 802.11, after a previous packet, the contention phase begins after 34uS long inter-frame space called the DIFS. In 802.11e this is called the AIFS and the length depends on the priority of the traffic intended for transmission. A station must wait a random back-off in terms of slots, which are 9 uS long is taken before every new transmission. The initial maximum number of slots depends on the traffic priority. In the 802.11a, there are 31 slots, while in 802.11e there are 15 slots of best-effort traffic, and even fewer or higher priority traffic like voice and video. If another station transmits, all other stations wait, and decrementing their back-off counter.

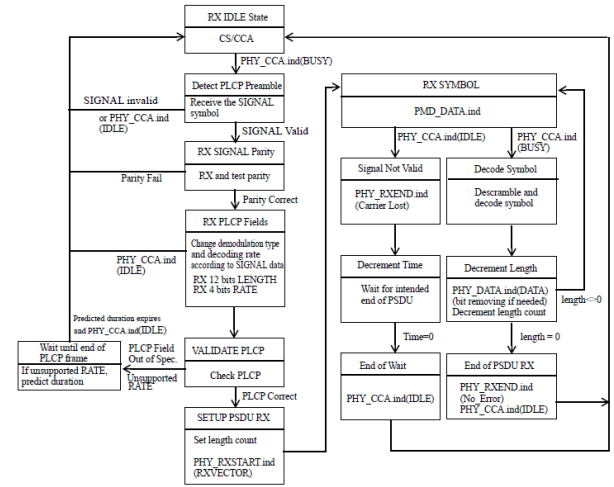

Figure 1. Receiver state machine [17]. If the VCS fails due to parity check failure, returns to IDLE state.

If two stations or more have the same back-off value there will be a collision. Listening stations will receive preamble from both STAs, which will appear similar to multi-path if such OFDM symbols fall within the cyclic prefix. Even if the preamble overlaps outside the cyclic prefix packet detection will likely be triggered. Given N colliding packets, each listening station may be in one of three states:

$S_1$: no packet can be decoded due to failure to pass parity check
$S_2$: a packet is incorrectly decoded but passes the parity check
$S_3$: one of the N packets is correctly decoded.

A station with MAC state $S_1$ will return to the idle state, according to Figure 1. A station with MAC in state $S_2$ will have its NAV set to a random value. Note this is largely fixed in 802.11n by using a CRC rather than a parity check in the HT header. A station in MAC state $S_3$, sets its NAV to cover the payload and the ACK. Clearly a station's NAV may be different from another station depending on the packet it decoded. Given a single cell, the undesirable consequences are mitigated by the ED mechanism.

*Proposition 1:* Even in non-fading channels where the VCS ought to work ideally, energy detection serves as a useful backup to the VCS.

In all S1-3, ED prevents additional collisions by holding the channel busy until overlapped packets are completed according to the standard "The HT PHY shall maintain PHY CCA.indication (BUSY , channel-list) until the received level drops below the CCA sensitivity level (for a missed preamble) specified in 20.3.21.5." [1].

In the above sensing operates close to ideal, and Bianchi's Markov chain analysis predicts slow loss in efficiency due to collisions as the number of STAs increases [11]. The "robustness" CSMA and has been demonstrated in a "large user scenario" by an experiment of a single Wi-Fi AP and many STAs in a single room, e.g. [18]. Moreover Bianchi's results compare favorably to NS-3 predictions, as shown in Section III.

*B. Overlapped packets and NS-3 modeling.*

NS-3 [20] is a state of the art event driven network simulator, modeling the complete protocol stack from application to physical layer. The most accurate but complex solution to handle overlapped packets is to simulate fully the functions of the receiver, i.e. packet detection, channel estimation, turbo decoding, and so forth as in [21]. NS-3 calculates the SINR over the length of the packet, and based on the packet-error-rate to SNR curves, obtained from physical layer simulator, determines whether packet detection was a success. Inspection of the modules and by traces shows that the Wi-Fi receiver model departs from Figure 1, the PLCP receive state machine. If two packets are generated simultaneously a receiver simply captures the first one in the queue, computing the SIR in the presence of the other.

We have introduced two changes to improve the existing abstraction. In the NS-3 file yans-wifi-phy.cc, packet capture is delayed until the end of the preamble which is 4 uS. At this point a new event called sync2packet is created for the strongest packet. This event is triggered at the end of the packet header, i.e. in 20 uS. The event handler of sync2packet computes the SINR over the packet header. If it is greater than 4 dB [22] the upper MAC is informed. It was also necessary to modify file interference-helper.cc to record all the overlapping events rather than collapsing the power as it does normally. If the SINR is less than 4 dB, the receiver returns to the IDLE state.

The rate manager is an important part of the link layer, selecting the MCS based on the success rate of previously sent packets. NS-3 includes a module modeling the Minstrel rate manager, which was shown experimentally to provide higher TCP rates than other rate managers [23]. A serious bug in NS-3's implementation was corrected [9]. Other parameters which control the operation of NS-3 are detailed in Table 1.

### III. ISOLATED SMALL NETWORK

Let us assume a small network with a few co-channel APs and evaluate the effect of loss in sensing on the protocol and on cell throughput. Cells and their STAs are placed according to an inter-cell distance, Figure 2. Intra-cell pathlosses are the pathlosses from one station to another station in the same cell. Inter-cell pathlosses are pathlosses from AP to AP, i.e. from cell center to center. They are a rough measure of how close or distant the cells and their respective stations are from each other. When the inter-cell pathloss is high enough packet headers are not detected or decoded incorrectly, and the cells operate independently, and their transmission are treated as fluctuating background noise.

| Parameter | Value |
|---|---|
| Rate | 802.11a OFDM 6 to 54 Mbps by Minstrel Rate Manager with update interval of 100 ms |
| MAC Protocol | Basic 802.11 access (no RTS/CTS) |
| Packet Capture Model | Custom SINR based |
| Packet Capture Threshold | -93 dBm AWGN channel |
| Packet Header | 68 bytes, including IP & UDP headers & padding |
| Packet Length | 1800 byte payload, 622 uS airtime |
| Rx Sensitivity | 2 RX Antenna, Atheros Enterprise Chipset |
| Channel | Rayleigh flat faded with 10 Hz Jakes Doppler (2 kph) |
| Transmit Power | +14 dBm fixed |
| Noise Floor | -94 dBm, 7 dB NF |
| Traffic Model | Full buffer via saturated UDP flows |
| Scheduler | Round Robin |

Table 1: Wi-Fi related parameters setting for NS-3.

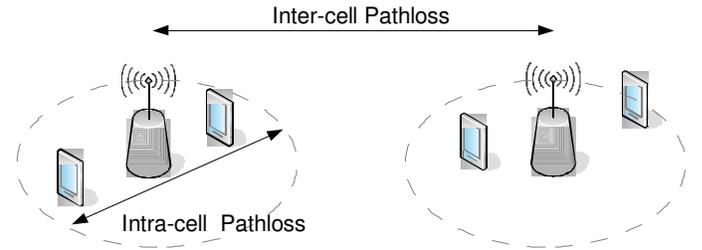

Figure 2: Two separated co-channels cells.

*A. Protocol & Sensing*

*Proposition 2:* Sensing failure due to collisions, fading or capture, triggers synchronization loss and permitting overlapped transmissions between cells.

*Illustration I: Collision*

Assume a topology similar to Figure 2, but with three APs. The inter-cell pathloss is moderate for functioning VCS but signals are below ED threshold. Each AP has data to transmit and is synchronized by a DIFS period by sensing the channel. Consider the timeline in Figure 3 with the events:

e1. Having identical back-off values $AP_1$ and $AP_2$ transmit together.
e2. $AP_3$ is unable to decode the headers of either 1 or 2 due to interference and starts later.
e3. $AP_1$ & $AP_2$ complete before $AP_3$ and after back-off $AP_1$ transmits.
e4. $AP_3$ isn't aware $AP_1$ is transmitting, so transmits.

While a packet collision occurs at $e_1$, reception may be successful depending on the STA SINR and the transmitted MCS. At AP 3 the overlapped packets are received at approximately zero dB SINR the VCS is unable to set the NAV. Due to failing to sense the channel busy, AP 3 transmits at $e_2$, followed by further un-synchronized packets $e_3$ and $e_4$. At $e_3$ when AP 1 listens it doesn't receive any packet preambles, and thus transmits concurrently with AP 3. The same occurs at $e_4$, when AP 3 finishes transmission, listens, but doesn't detect any packet preambles, and thus transmits.

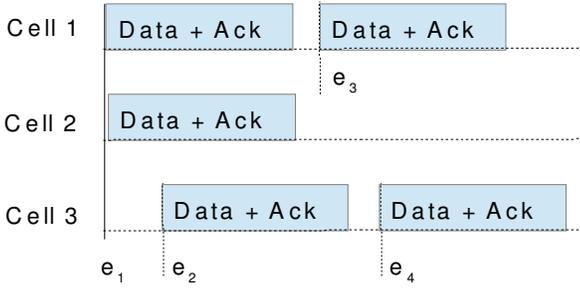

Figure 3: Example of timeline (left to right) of transmissions in three cells. The term "Data" represents the entire 802.11 preamble, header, and MPDU.

*Illustration II: Fading*

Assume two APs initially synchronized to a DIFS, but short term fading causes VCS at $AP_2$ to miss $AP_1$'s packet header. In Figure 4, although they have different back-off values, both $AP_1$ and $AP_2$ will transmit, at $e_1$ and $e_2$ respectively. Later during packet transmission, assume the fading event is finished and VCS sensing at both $AP_1$ and $AP_2$ is operational. As shown in the figure due to the different packet lengths and starting times, $AP_1$ begins the second DIFS period before $AP_2$, and after back-off transmits before $AP_2$ completes its transmission. Now $AP_2$ waits for DIFS and back-off, detects no preambles and transmits.

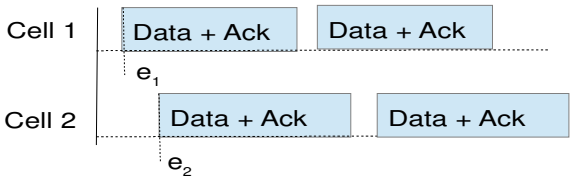

Figure 4: Time line of transmissions in two cells. Fading causes initial sync loss and overlapped transmission, followed by continued overlap due to sensing misalignment.

### B. Throughput & Sensing

Cell throughput is computed for scenarios with 2 and 4 APs as a function of the number of users, and inter-cell pathloss. Propagation channels are time varying with a Rayleigh distribution. The average pathloss between cells is fixed at the same value for all AP/STAs. The average intra-cell pathloss is fixed at 64 dB so that both VCS and ED are active. The MCS is fixed to 24 Mbps. Uplink UDP traffic is generated, such that there is always a full buffer. NS-3 is configured to report the number of received packets at completion of the simulation period.

Figure 5 and 6 present the throughput results for varying number of users and inter-cell pathlosses. Single AP throughput is provided for reference as indicated by the blue circle. We observe:

- *Cell total throughput is limited to single cell AP when the pathloss is such that the ED is active.* See Figure with PL=64 dB where the out of cell RSSI is -50 dBm, as the throughput curves overlap. The MAC operates nearly ideally since as described by Proposition 1 the ED mechanism quickly resynchronizes all the STAs when collisions or fading causes synchronization loss.

- *Cell total throughput grows with increase separation when the ED is not active.* Observed in curves where PL is 86, 96, 106 dB. The ED threshold is not met by the out-of-cell transmissions which are received at -72, -82, and -92 dBm. respectively. This is 1 dB above the minimum required to decode the packet header so the VCS is effective in the AWGN channel. Since probability of VCS failure increases with average pathloss, and VCS failure according to Propositions 2a and 2b leads to overlapped transmissions, and these increases throughput being from separated cells.

- *Initial rise in throughput vs. number of users.* This behavior may be attributed to increase in collisions that cause overlapped transmissions according to Proposition 2.

- *Subsequent decay in throughput vs. number of users.* This is due to in cell collisions.

- *Decay in throughput falls according to Bianchi's model for single cell & when ED is active.* Note the throughput was normalized at single user to give consistent amount of overhead. NS-3 results closely follow Bianchi's model, thus validating NS-3's CSMA model in ideal single cell scenarios.

## IV. MULTI-CELL CAPACITY SCALING WITH DENSITY

The throughput of a large frequency planned network is computed at various densities from isolated to overlapping.

### A. Experiment Design

APs are positioned on a square grid, with coordinate wrap around to eliminate edge effects, i.e. the bottom cells see interference from the top. Figure 7 shows the user locations (STAs) around the APs and the reuse plan 12. To simplify the presentation, traffic is full-buffer which may roughly correspond to a scenario of FTP download to each of the stations. The APs and STAs transmit at fixed RF power

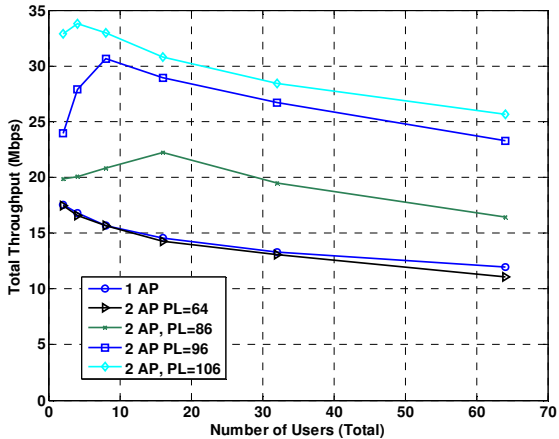

Figure 5. Two cell throughput vs. number of users for different values of inter-cell pathloss.

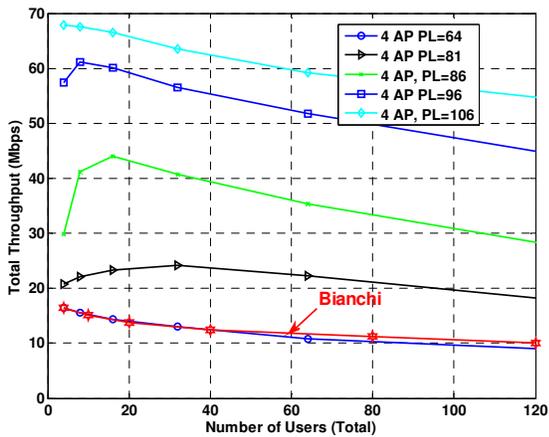

Figure 6. Four cell throughput vs. total number of users for different values of inter-cell pathloss.

Frequency reuse puts certain amount of distance between the co-channel cells, reducing interference but reducing the bandwidth available per cell. An aggressive reuse plan with 4 channel groups (reuse 4) and less aggressive plan with 12 channels (reuse 12) is simulated. While the actual number of 5 GH unlicensed band channels varies with local regulation, a 12 channels for total of 240 MHz bandwidth is assumed here. With reuse 4 each AP has 3 separate 20 MHz channels, whereas for reuse 12 each AP only has one 20 MHz channel. Only the co-channel cells are simulated, and the nearest neighbor is every other cell for reuse 4. A total 36 of APs and 4 STAs per AP, for total of 144 STAs were simulated.

Indoor pathloss at distance $r$ from the source is given by [5]:

$$P_G = \frac{\lambda^2}{4\pi}\left[\frac{\kappa_d}{4\pi r} + \frac{1}{4\pi r^2}\right]e^{-\kappa_d r} \quad (1)$$

where $\lambda$ is the wavelength in meters, and $\kappa_d$ is the absorption coefficient taken. Using (1) with $\kappa_d = 0.24$, along with lognormal shading fading of 4 dB, propagation is computed between all STAs and APs. Fast fading due to mobility is a random process according to the Jakes Doppler spectrum.

The ED range is 15 meters for the average pathloss according to (1) and no fading.

*B. Results & Analysis*

The relative capacity providing efficiency $E$ of an AP is defined as fractional change in area capacity divided by the fractional change in cell density. $E$ is also the area capacity $C_1$ at new density $D_1$ divided by the area capacity at the original density, i.e.

$$E = \frac{C_1}{D_1} / \frac{C_0}{D_0}, \quad (2)$$

and $E=1$ means that a new cell brings 100% of the throughput at the original density.

The basic data, i.e. user throughputs, from which scaling is derived is shown in Figure 8 for reuse 12. The CDFs illustrates the effect of shrinking ISD on throughputs and fairness. The worsening of these key metrics is due to decrease in SINR. At 40 m ISD where the cells should be operating ideally the throughput ratio between worse users (10%) and the best (90%) is 1.3. At 10 m ISD this ratio is much worse at 2.6.

Area capacity is plotted as a function of relative cell density in Figure 9. The relative cell density is cell area divided by the reference cell area at 40 m ISD. Initially the area capacity of aggressive reuse (reuse 4) is much higher than less aggressive (reuse 12), but with densification, the capacity saturates quicker and the advantage diminishes. Initially increasing cell density provides full area capacity gains with $E \approx 1$. As the cells begin to overlap, the $E$ decreases to about 0.65, i.e. for every new cell added the capacity increases by 65% as it would in the low density regime.

The area capacity trend is driven by two effects: MAC interaction and SINR degradation. To separate the effects we compare Wi-Fi area capacity to the capacity of regular licensed band FDD LTE. FDD LTE downlink may transmit continuously in reuse 1 fashion. This is not to be confused with LTE-U or LAA which has a listen-before-talk feature for use in the unlicensed band. Using the downlink SINRs obtained as if all APs were broadcasting at full power, a rate mapper is applied to compute the "LTE" rates. For comparison the LTE system is only given one 20 MHz channel vs. the 240 MHz given to the Wi-Fi network. Despite 1/12 the bandwidth the initial LTE area capacity is actually about the same as 802.11a Wi-Fi reuse 4. Note that 802.11n due to MIMO and aggregation will have higher starting point. While the absolute throughput values will vary as both LTE and Wi-Fi evolve, results show LTE area capacity also reduced due to interference, with $E$ about 0.8.

CONCLUSIONS

The VCS and ED mechanisms jointly prevent in-cell collisions, but for separated cells the VCS permits cells to lose synchronization. For the densities tested Wi-Fi area capacity grows with densification but each new cell provides less capacity than a cell at the original density. This penalty is due to decreasing SINR and partial MAC coordination. Future work would focus on performance scaling for uplink and bi-directional TCP traffic.

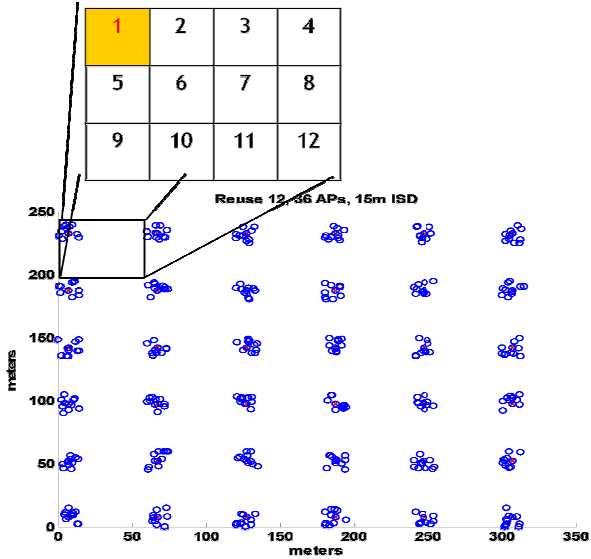

Figure 7: Square cell centers with frequency plan of 12. Red circles are the APs and blue circles are STA locations.

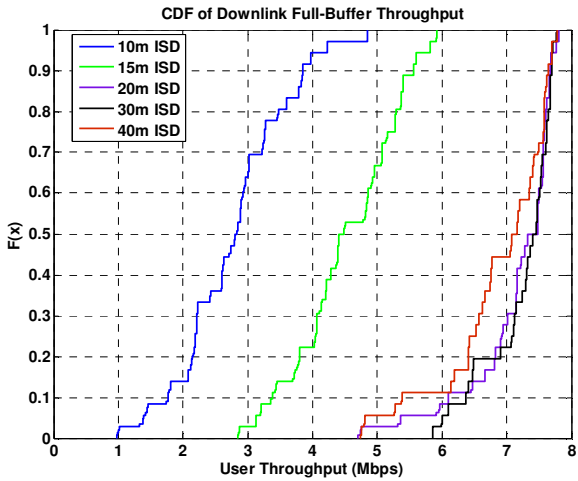

Figure 8: User throughputs at 10 to 40 m ISD and reuse 12.

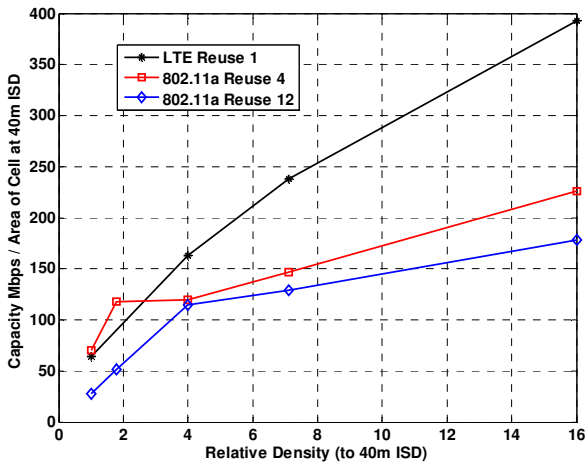

Figure 9: Downlink area capacity vs. density for Wi-Fi for reuse 4 and reuse 12, along with LTE at reuse 1.